\documentclass[amsmath,amssymb,10pt]{iopart}

\usepackage{cases}
\usepackage{graphicx}
\usepackage{dcolumn}
\usepackage{bm}
\begin{document}

\title[Electromagnetic coupling of twisted multi-filament superconducting tapes in a ramped magnetic field]
{Electromagnetic coupling of twisted multi-filament superconducting tapes in a ramped magnetic field}

\author{Yoichi Higashi}
\address{National Institute of Advanced Industrial Science and Technology (AIST), Tsukuba, Ibaraki 305-8568, Japan}
\ead{y.higashi@aist.go.jp}
\author{Yasunori Mawatari}
\address{National Institute of Advanced Industrial Science and Technology (AIST), Tsukuba, Ibaraki 305-8568, Japan}
\vspace{10pt}
\begin{indented}
\item[]\today
\end{indented}

\begin{abstract}

We investigate theoretically the magnetization loss and electromagnetic coupling of twisted multi-filament superconducting (SC) tapes in a ramped magnetic field. Based on the two-dimensional reduced Faraday--Maxwell equation for a tape surface obtained with a thin-sheet approximation, we simulate numerically the power loss $P$ per unit length on twisted multi-filament tapes in the steady state. The current density profile clearly shows electromagnetic coupling between the SC filaments upon increasing the field sweep rate $\beta$. Although the $\beta$ dependence of $P/\beta$ for twisted multi-filament SC tapes closely resembles that for filaments in an alternating field, we show that the mechanism for electromagnetic coupling in a ramped field differs from that in an alternating field. We also identify the conditions under which electromagnetic coupling is suppressed for the typical sweep rate of a magnet used for magnetic resonance imaging.
\end{abstract}

%
%
\vspace{2pc}
\ioptwocol
%

\section{Introduction}
Superconducting (SC) cables and wires based on rare-earth barium copper oxide (ReBCO) SC tape are being actively researched and developed, with a possible target application being magnets for nuclear magnetic resonance and magnetic resonance imaging (MRI). ReBCO SC tape exhibits high critical current density even in a high magnetic field. However, its tape-shaped geometry leads to considerable power being lost in response to the perpendicular component of a time-varying magnetic field. To reduce this loss for an SC tape, we must reduce the size of the loops of the current streamlines. Previous experimental and numerical work has shown that multi-filamentarization is an effective way to reduce losses associated with an SC tape in an alternating magnetic field \cite{amemiya2004, sumption2004, amemiya2006, grilli2016}.

In practical SC cables, the SC filaments are coated with a stabilizer (e.g., copper) to ensure that the SC tape is thermally stable. However, this leads to a coupling current flowing between the SC filaments via the stabilizer in an alternating magnetic field whose frequency is high.
Therefore, multi-filamentarization is ineffective for reducing the power loss at high frequencies. The coupling frequency $\omega_{\rm c}$ at which the SC filaments become electromagnetically coupled is known to be related to (i) the normal resistivity $\rho_{\rm n}$, which short-circuits the SC filaments, and (ii) the twist pitch length $L_{\rm p}$ as $\omega_{\rm c} \propto \rho_{\rm n}/L^2_{\rm p}$ \cite{wilson1983}, meaning that the electromagnetic coupling depends on $L_{\rm p}$. Thus, by shortening the effective tape length with twisting \cite{takayasu2012}, electromagnetic coupling can be suppressed.

This article reports on a theoretical study of the electromagnetic coupling of twisted multi-filament SC tapes exposed to a constantly ramped magnetic field on the supposition that an MRI magnet is magnetized/demagnetized by a transport current. Clarifying the conditions for electromagnetic coupling gives valuable information regarding how fast the magnetic field can be swept and how tightly the SC filaments should be twisted given the practical sweep rates used when operating an MRI magnet. Furthermore, we show that the mechanism for coupling SC filaments electromagnetically in a ramped field is essentially different from that in an alternating field.
\section{Model of twisted multi-filament superconducting tape}
\begin{figure}[t]
  \begin{center}
    \begin{tabular}{p{55mm}p{70mm}}
           \resizebox{55mm}{!}{\includegraphics{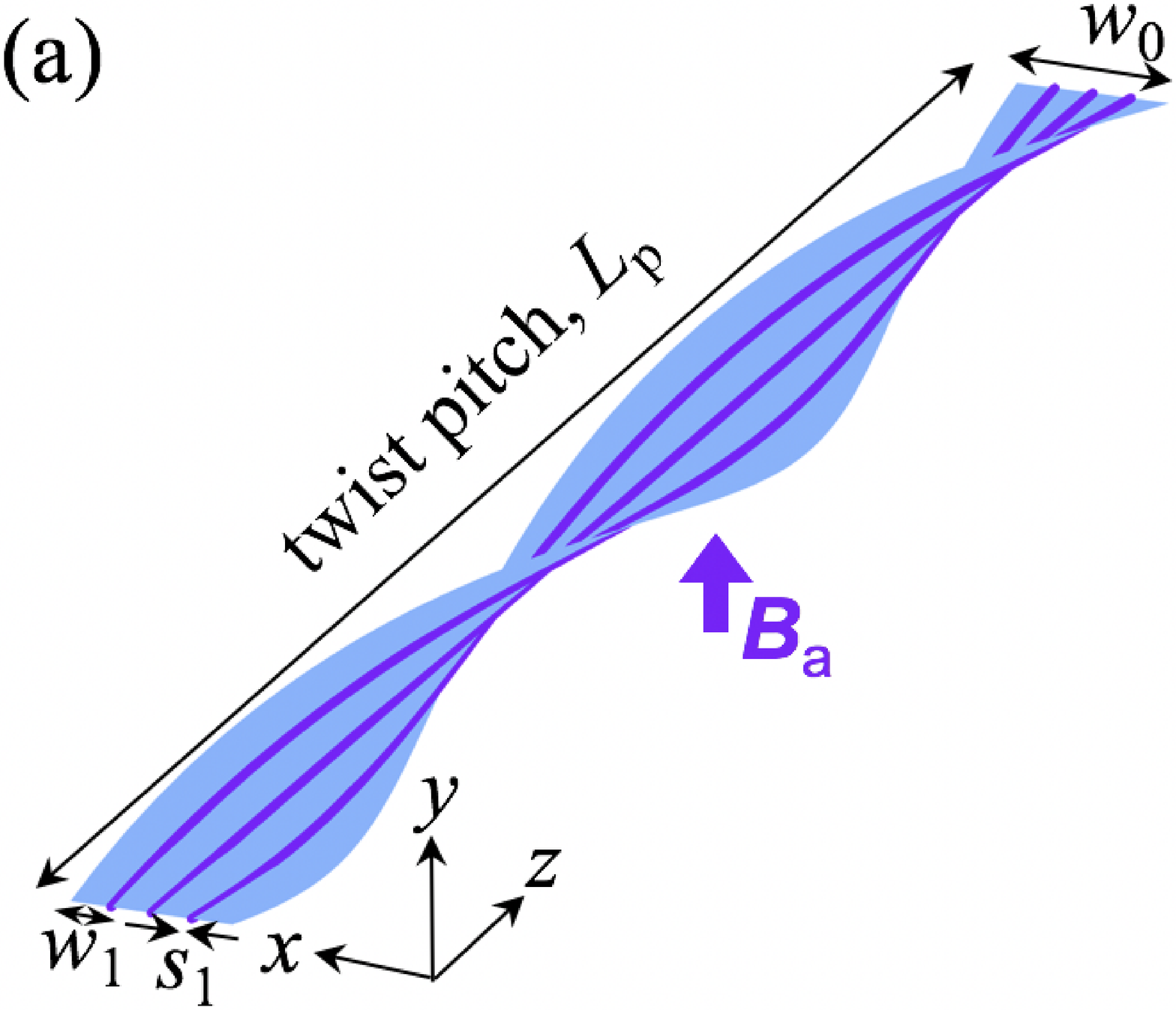}}\\
           \resizebox{70mm}{!}{\includegraphics{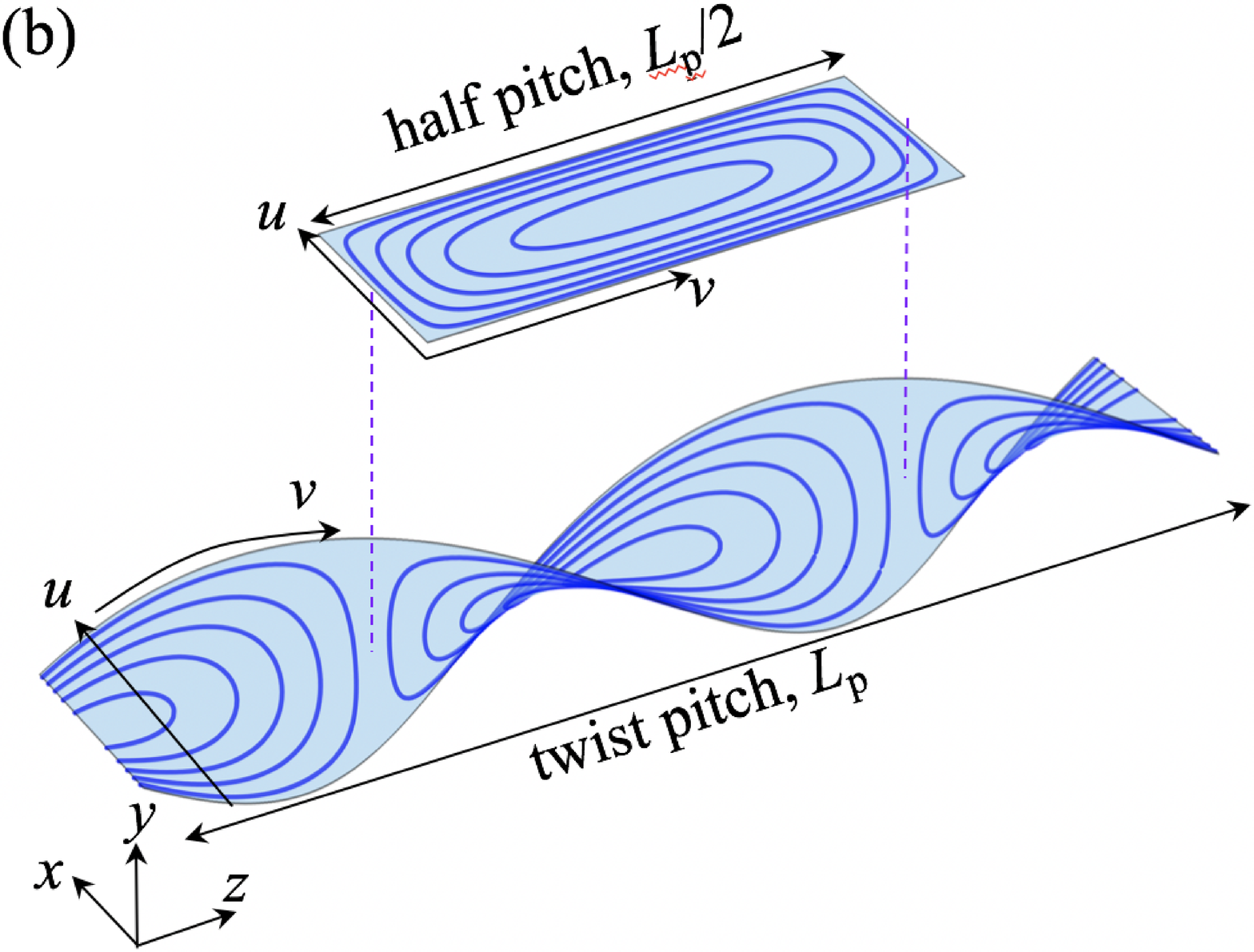}}\\
    \end{tabular}
\caption{
\label{fig1}
(a) Schematic of twisted four-filament superconducting (SC) tape with filament width $w_1$, total width $w_0$, and twisted pitch length $L_{\rm p}$. The thin dark lines depict resistive slots of width $s_1$. An external field $B_{\rm a}$ is applied in the $y$ direction. (b) Schematic of numerically analyzed area corresponding to half pitch $L_{\rm p}/2$ on surface of twisted SC tape. The axes $(u,v)$ are on the tape surface. The solid lines on the tape surface depict the current streamlines. The resistive slots are omitted here for clarity.
}
  \end{center}
\end{figure}
We consider a twisted four-filament SC tape with resistive slots between the filaments, as shown in Fig.~\ref{fig1}(a). The total tape width, SC filament width, and slot width are $w_0$, $w_1$, and $s_1$, respectively. The thickness $d_0$ of the SC filaments and resistive slots is a common thickness. In practice, the SC tape is coated with a normal metal stabilizer such as copper. Here, we model this normal metal stabilizer by means of resistive slots embedded between the SC filaments \cite{amemiya2018}. Because $d_0 \ll w_0$, a twisted multi-filament SC tape can be approximated as a twisted multi-filament tape surface of infinitesimal thickness [Fig.~\ref{fig1}(a)] that is described by the following coordinates:
\begin{equation}
\label{helicoid-coordinate}
\cases{
x=u \cos kv-\eta \sin kv,\\
y=\eta \cos kv + u \sin kv,\\
z=v,
}
\end{equation}
where $k=2\pi/L_{\rm p}$ with $L_{\rm p}$ the twist pitch length. The $u$ and $v$ axes are on the twisted tape surface as shown in Fig.~\ref{fig1}(b). With $\hat{\bm{u}}$ and $\hat{\bm{v}}$ denoting the respective unit vectors, note that $\hat{\bm{u}}\perp \hat{\bm{v}}$ on the tape surface at $\eta=0$.
The SC filaments correspond to $\eta=0$, $s_1/2<|u|<s_1/2+w_1$, $w_0/2-w_1<|u|<w_0/2$, and $-\infty<v<\infty$. The resistive slots correspond to $\eta=0$, $|u|\le s_1/2$, $s_1/2+w_1 \le |u| \le w_0/2-w_1$, and $-\infty < v<\infty$.

\section{Electromagnetic response of twisted multi-filament superconducting tape}
The present interest is in losses in the case of excitation/demagnetization of an MRI magnet. In a high magnetic field, the transport loss is negligible compared to the magnetization loss \cite{kajikawa2016}; therefore, we focus on the magnetization loss of a twisted multi-filament SC tape in a swept external magnetic field. Note that the presence of a transport current reduces the magnetization loss \cite{leblanc1963}. Furthermore, we assume the steady state in which the strength of the external magnetic field exceeds that required for full flux penetration.

We begin with Faraday's law of induction in the steady state with the magnetic field parallel to the $y$ axis:
\begin{eqnarray}
\bm{\nabla} \times \bm{E}=-\frac{\partial \bm{B}}{\partial t}\approx-\beta\hat{\bm{y}},
\label{Faraday}
\end{eqnarray}
where $\beta$ is the field sweep rate, $\bm{E}$ and $\bm{B}$ are the electric field and the magnetic induction, respectively, and $\hat{\bm{y}}$ is the unit vector in the $y$ direction. The magnetic field due to the screening current is neglected in Eq.~(\ref{Faraday}).
We also take into account (i) the thin-sheet approximation \cite{zhang2017} and (ii) the response to the perpendicular component of the external magnetic field. Consequently, the Faraday--Maxwell equation on a twisted multi-filament SC tape is reduced to
\begin{eqnarray}
\frac{\partial}{\partial u}\left[ \rho \frac{\partial g}{\partial u}(1+k^2u^2) \right]+\frac{\partial}{\partial v}\left[\rho \frac{\partial g}{\partial v} \right]=\beta \cos kv,
\label{poisson}
\end{eqnarray}
where $g(u,v)$ is a scalar function that describes the current streamlines.
The resistivity $\rho$ of a multi-filament tape with resistive slots is described by
\begin{eqnarray}
\rho=
\label{cases}
\cases{
\frac{E_{\rm c}}{J_{\rm c}}\left( \frac{|\bm{J}|}{J_{\rm c}} \right)^{n-1}\equiv \rho_{\rm sc}(|\bm{J}|)~~{\rm for~SC~filaments},\\
\rho_{\rm n}~~{\rm for~resistive~slots},
}
\label{n-value model}
\end{eqnarray}
which is defined via $\bm{E}=\rho\bm{J}$. Here, $\bm{J}$ is the current density, $J_{\rm c}$ is the critical current density, and $E_{\rm c}$ is the electric field criterion. The derivation of Eq.~(\ref{poisson}) is described in detail in Ref.~\cite{higashi2018}.

\section{Current density profiles on multi-filament twisted tape}
\begin{figure*}[h]
  \begin{center}
    \begin{tabular}{p{164mm}}
            \resizebox{164mm}{!}{\includegraphics{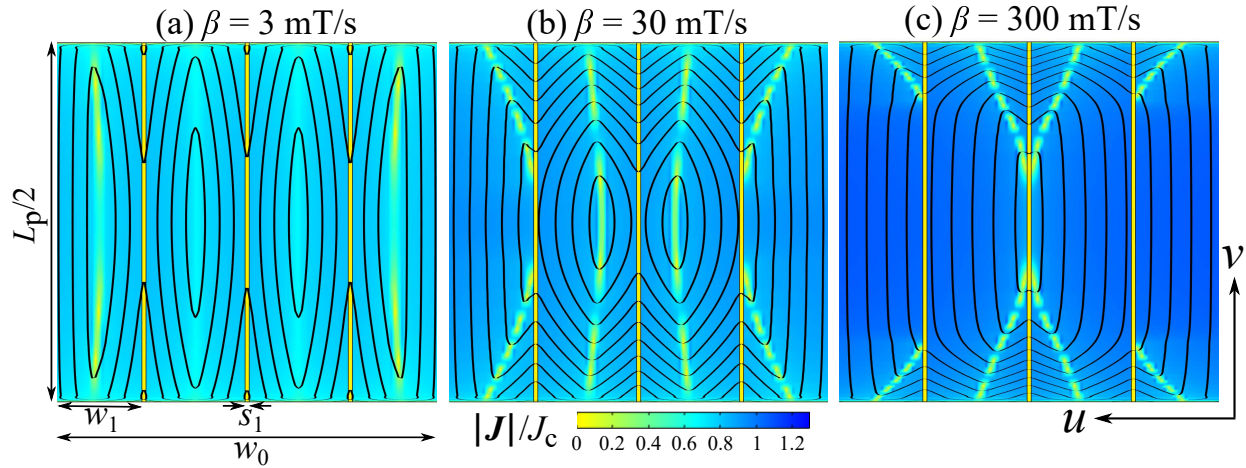}}
    \end{tabular}
\caption{
\label{fig2}
Current density profiles on a twisted multi-filament tape with a half pitch for a field sweep rate $\beta$ of (a) 3~mT/s, (b) 30~mT/s, and (c) 300~mT/s. The twist pitch length is set to $L_{\rm p}=500w_0=2$~m. The horizontal and vertical directions correspond to the $u$ and $v$ axes, respectively. The solid black lines  depict the current streamlines [i.e., the contour lines of $g(u,v)$].
}
  \end{center}
\end{figure*}
We obtain the spatial profile of the electric current density on the tape surface through the scalar function $g(u,v)$ that describes the current streamlines \cite{higashi2018} [see Fig.~\ref{fig1}(b)]:
\begin{equation}
J_u=-\frac{\partial g}{\partial v},~~J_v=\frac{\partial g}{\partial u}\sqrt{1+k^2u^2}.
\end{equation}
In this study, $g(u,v)$ is obtained by solving Eq.~(\ref{poisson}) numerically using the commercial software COMSOL Multiphysics\textsuperscript{\textregistered} \cite{comsol}. The dimensions of the twisted multi-filament SC tape are set to $w_0=4$~mm, $w_1=0.97$~mm, $s_1=40$~$\mu$m, $d_0=2$~$\mu$m, and $L_{\rm p}=2$~m. The parameters of the multi-filament SC tape with resistive slots are set to $J_{\rm c}=5\times 10^{10}$~A/m$^2$, $E_{\rm c}=1$~$\mu$V/cm, $n=21$, and $\rho_{\rm n}=1\times 10^{-9}$~$\Omega$m \cite{amemiya2018}.

Note that twisting the tape surface divides the current streamlines into closed loops within every half pitch $L_{\rm p}/2$, as shown in Fig.~\ref{fig1}(b). Indeed, we confirmed that no current crosses the boundary at $v=L_{\rm p}/4$ between the current loops, that is, $J_v(u,v=L_{\rm p}/4)\approx 0$. Thus, one may consider a twisted tape with a half pitch instead of one with a full pitch. Therefore, we impose Dirichlet boundary conditions at the long edges, that is, $g(u=\pm w_0/2,v)=0$, and where the twisted tape surface is parallel to the magnetic field ($||~\hat{\bm{y}}$), that is, $g(u,v=\pm L_{\rm p}/4)=0$.

Figure~\ref{fig2} shows spatial profiles of the electric current density normalized by $J_{\rm c}$ (i.e., $|\bm{J}(u,v)|/J_{\rm c}$) on a multi-filament twisted SC tape with half pitch length $L_{\rm p}/2$. The solid black lines in Fig.~\ref{fig2} depict the current streamlines expressed by the contour lines of $g(u,v)$. For two filaments in an alternating magnetic field, similar current density and streamline profiles have been obtained previously based on the variational principle and the finite element method \cite{pardo2016}. At $\beta=3$~mT/s [Fig.~\ref{fig2}(a)], most of the loops of the current streamlines are closed within each SC filament, making multi-filamentarization effective for reducing the loss. At $\beta=30$~mT/s [Fig.~\ref{fig2}(b)], several current streamlines cross the resistive slots. Thus, as seen in Fig.~\ref{fig3}(a), the coupling loss exhibits a maximum for the present choice of parameters. At $\beta=300$~mT/s [Fig.~\ref{fig2}(c)], the loops of the current streamlines spread across the entire width of the multi-filament twisted tape, making the filaments behave collectively as a single twisted tape. In this case, the twisted multi-filament SC tape is completely electromagnetically coupled, making multi-filamentarization ineffective for reducing the loss. To summarize, the filaments are electromagnetically decoupled for $\beta \ll \beta_{\rm c}=30$~mT/s but coupled for $\beta>\beta_{\rm c}=30$~mT/s.

\section{Electromagnetic coupling of multi-filament twisted tape}
\begin{figure*}[tb]
  \begin{center}
    \begin{tabular}{p{82mm}p{82mm}}
            \resizebox{82mm}{!}{\includegraphics{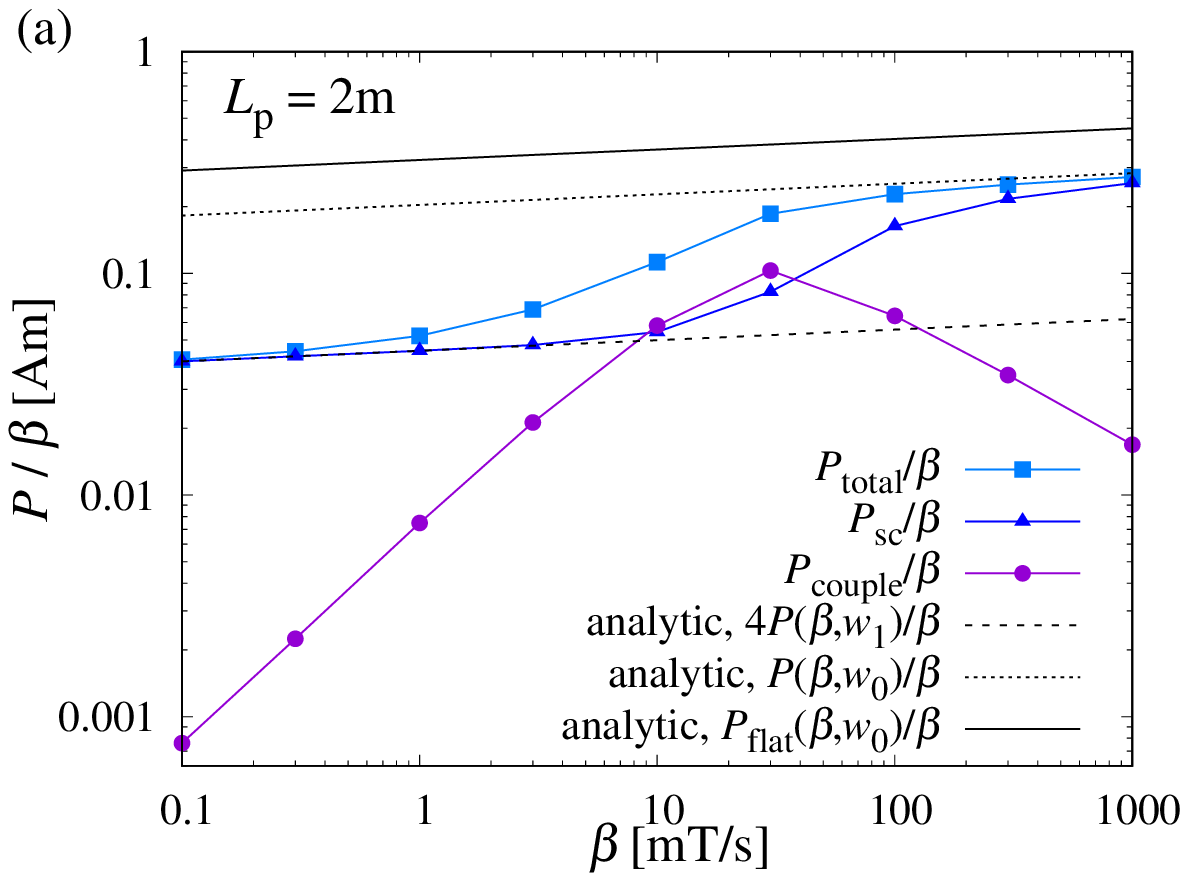}}&
            \resizebox{82mm}{!}{\includegraphics{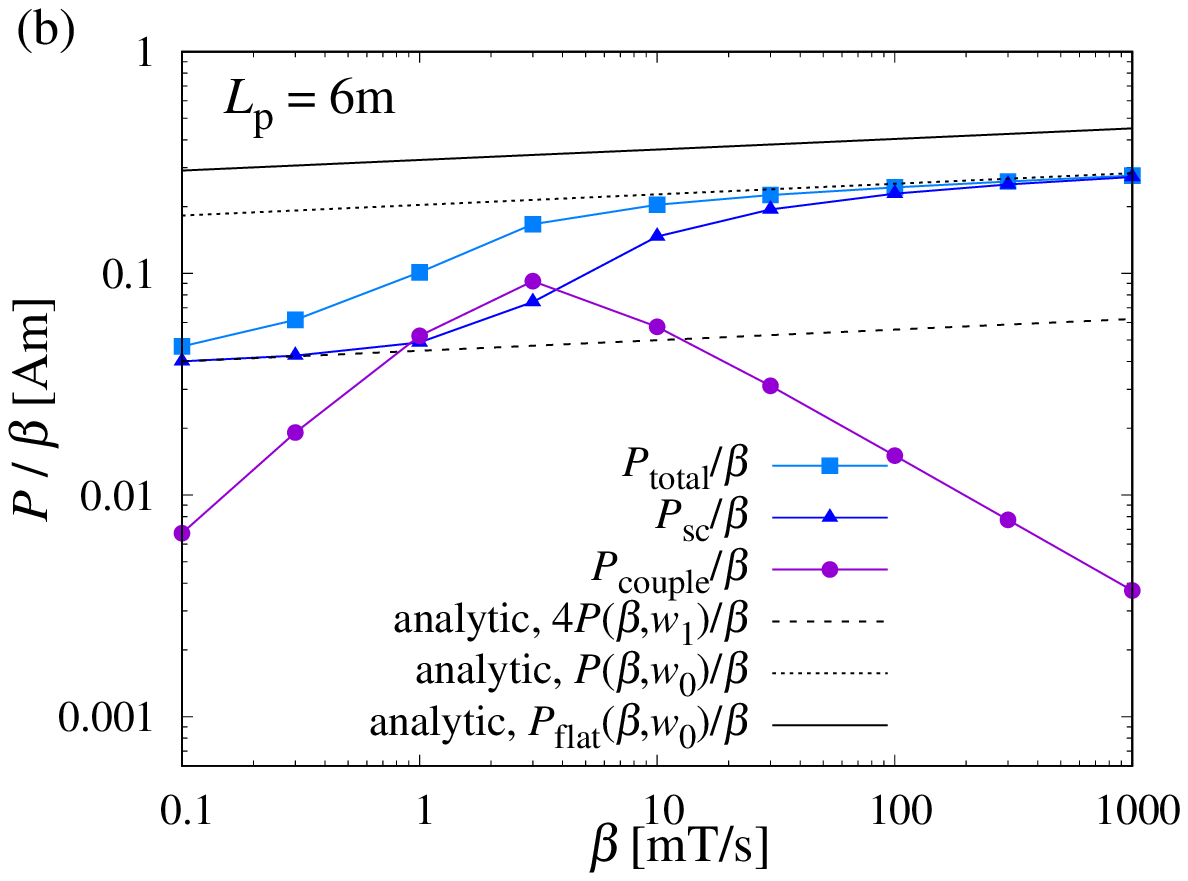}}
    \end{tabular}
\caption{
\label{fig3}
Dependence of power loss $P$ per unit of length divided by sweep rate $\beta$ on $\beta$ for $L_{\rm p}=$ (a) 2~m and (b) 6~m. The total power loss, that of the SC filaments, and that of the resistive slots are denoted by squares, triangles, and circles, respectively.
The solid, dotted, and dashed lines are those evaluated analytically via 
$P_{\rm flat}(\beta,w_0)/\beta$ [Eq.~(\ref{loss-flat})], $P(\beta,w_0)/\beta$ [Eq.~(\ref{loss-power-per-beta})], and $4P(\beta,w_1)/\beta$ [Eq.~(\ref{loss-power-per-beta})], respectively.
}
  \end{center}
\end{figure*}
In the limit $L_{\rm p}\rightarrow \infty$ for a loosely twisted tape, the power loss per unit length has been obtained analytically \cite{higashi2018}. For a twisted SC tape, the hysteresis power loss $P$ per unit length divided by $\beta$ is recast as
\begin{eqnarray}
{P(\beta,w)\over \beta}=\frac{B\left({2n+1 \over 2n},{1 \over 2} \right)}{\pi} \frac{P_{\rm flat}(\beta,w)}{\beta},
\label{loss-power-per-beta}\\
P_{\rm flat}(\beta,w)=\left( \frac{\beta w}{2 E_{\rm c}} \right)^{1+1/n}\frac{E_{\rm c}J_{\rm c}wd_0}{2+1/n},
\label{loss-flat}
\end{eqnarray}
where $B(p,q)=2\int_0^{\pi/2}{\rm d}\theta \cos^{2p-1}\theta \sin^{2q-1}\theta$ is the beta function with positive real numbers $p$ and $q$, $w$ is the effective tape width, and $P_{\rm flat}$ is the power loss per unit length for a flat tape. In the Bean limit of $n\rightarrow \infty$, we have $P/\beta \sim J_{\rm c}d_0w^2$. In the case of $N$ filaments, we have $w_1\equiv w_0/N$ and $P/\beta \sim J_{\rm c}d_0w^2_1 N=J_{\rm c}d_0w^2_0/N$. Therefore, $P$ is reduced to $1/N$ times that for a non-striated tape with no electromagnetic coupling. Herein, we fix the number of SC filaments to $N=4$.


The power losses per unit length for the SC filaments ($P_{\rm sc}$), the normal resistive slots ($P_{\rm couple}$), and the entire twisted multi-filament tape ($P_{\rm total}$) are evaluated numerically via
\begin{eqnarray}
&P_{\rm sc}=\int\int_{\rm filaments}{\rm d}u{\rm d}v p(u,v),\\
&P_{\rm couple}=\int\int_{\rm slots}{\rm d}u{\rm d}v p(u,v),\\
&p(u,v)=\frac{d_0}{L_{\rm p}/2}(E_uJ_u+E_vJ_v),\label{coupling-loss-density}\\
&P_{\rm total}=P_{\rm sc}+P_{\rm couple}
\end{eqnarray}
(see Appendix).
Figure~\ref{fig3}(a) and (b) show how the power loss $P$ per unit length divided by $\beta$ depends on $\beta$ for $L_{\rm p}=2$~m and 6~m, respectively. Therein, the solid, dotted, and dashed lines correspond to the power losses for a flat tape, a twisted tape, and twisted multi-filament tape as evaluated analytically from $P_{\rm flat}(\beta,w_0)/\beta$, $P(\beta,w_0)/\beta$, and $4P(\beta,w_1)/\beta$, respectively. Note that the effect of the SC filaments being wound around the axis of the helicoid at $u=0$ is not present in $4P(\beta,w_1)/\beta$; however, it turns out not to be crucial  because $4P(\beta,w_1)/\beta$ agrees quantitatively with $P_{\rm sc}/\beta$ at low values of $\beta$.

In Fig.~\ref{fig3}(a), $P_{\rm couple}/\beta$ (circles) has a maximum at $\beta_{\rm c}\approx 30$~mT/s. Herein, we refer to $\beta_{\rm c}$ as the coupling sweep rate. Meanwhile, with increasing $\beta$, $P_{\rm sc}/\beta$ (triangles) increases gradually from the value expected theoretically for electromagnetically decoupled SC filaments, that is, $4P(\beta,w_1)/\beta$ (dashed line), to that for completely coupled filaments, that is, $P(\beta,w_0)/\beta$ (dotted line).
Meanwhile, $P_{\rm total}/\beta$ (squares) increases toward $P(\beta,w_0)/\beta$ with increasing $\beta$. Note that $P_{\rm total}/\beta$ almost reaches $P(\beta,w_0)/\beta$ at $\beta \approx \beta_{\rm c}$.

At $\beta=1,000$~mT/s ($\gg\beta_{\rm c}$), the SC filaments are completely coupled electromagnetically and behave collectively as a single SC tape with no multi-filamentarization. The numerical values of $P_{\rm sc}/\beta$ are fitted well by those evaluated analytically via $P(\beta,w_0)/\beta$ [Eq.~(\ref{loss-power-per-beta}), dotted line in Fig.~\ref{fig3}]. Thus, multi-filamentarization is ineffective for reducing the magnetization loss in the case of $\beta \gg \beta_{\rm c}$. By contrast, at $\beta=0.1$~mT/s ($\ll \beta_{\rm c}$), the SC filaments are completely decoupled electromagnetically. Now, $P_{\rm sc}/\beta$ is fitted well by $4P(\beta,w_1)/\beta$ [dashed line in Fig.~\ref{fig3}] and multi-filamentarization is effective for reducing the hysteresis loss to a quarter of that for a non-striated twisted tape (note that $N=4$).

In the case of $L_{\rm p}=6$~m [Fig.~\ref{fig3}(b)], $\beta_{\rm c}$ is less than that for $L_{\rm p}=2$~m by an order of magnitude, that is, $\beta_{\rm c}=3$~mT/s, because $\beta_{\rm c}\propto1/L^2_{\rm p}$ \cite{wilson1983}. Therefore, at the typical sweep rate $\beta_{\rm MRI}\sim1$~mT/s for an MRI magnet \cite{yokoyama2017,yachida2017}, which is the same order of $\beta_{\rm c}$ as that for $L_{\rm p}=6$~m, the effect of multi-filamentarization is restricted to reduce the total loss. However, for $L_{\rm p}=2$~m, multi-filamentarization is effective at reducing the total loss for $\beta_{\rm MRI} \ll \beta_{\rm c}$.

\section{Coupling loss}
\begin{figure}[t]
  \begin{center}
    \begin{tabular}{p{80mm}}
            \resizebox{80mm}{!}{\includegraphics{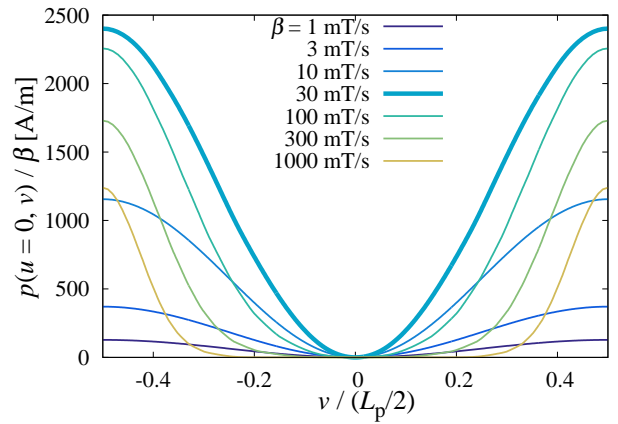}}
    \end{tabular}
\caption{
\label{fig4}
Spatial profile of coupling-loss power density for $L_{\rm p}=$2~m along the slot at $u=0$.
The field sweep rate $\beta$ is varied from 1~mT/s to 1,000~mT/s. The profile at $\beta_{\rm c}=30$~mT/s is indicated with the bold line.
}
  \end{center}
\end{figure}
\begin{figure}[t]
  \begin{center}
    \begin{tabular}{p{80mm}}
            \resizebox{80mm}{!}{\includegraphics{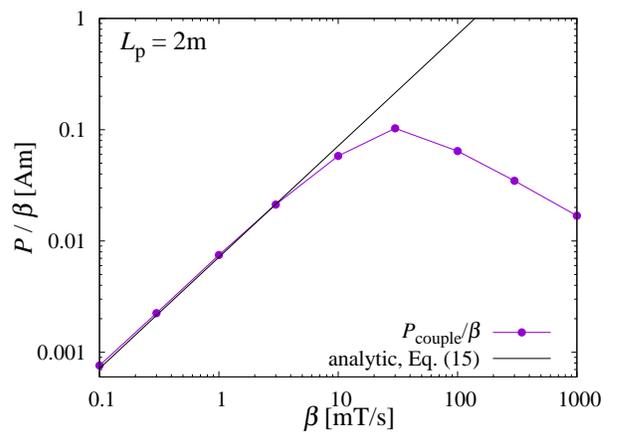}}
    \end{tabular}
\caption{
\label{fig5}
Dependence of coupling-loss power per field sweep rate $\beta$ on $\beta$ for $L_{\rm p}=2$~m fitted by Eq.~(\ref{coupling-loss-low-sweep-rates}).
}
  \end{center}
\end{figure}
Figure~\ref{fig4}
shows the spatial profiles of the coupling-loss power density $p(v)/\beta$ [Eq.~(\ref{coupling-loss-density})] along the resistive slot at $u=0$. In the neighboring slot at $u=w_1+s_1$, the spatial profile and the dependence of $p(v)$ on $\beta$ are qualitatively the same as those along the slot at $u=0$. 
For $\beta \le \beta_{\rm c}=30$~mT ($L_{\rm p}=2$~m), the spatial expansion of $p(v)/\beta$ is almost invariant, and the magnitude of $p(v)/\beta$ increases with increasing $\beta$. However, for $\beta > \beta_{\rm c}$ and with increasing $\beta$, the spatial expansion of $p(v)/\beta$ as well as its amplitude decreases, resulting in reduced coupling loss at large values of $\beta$. The shrinkage of the spatial expansion of $p(v)/\beta$ at large $\beta$ is due to (i) the local increase of the transverse current density $J_u$ at $v=\pm L_{\rm p}/4$ and (ii) the localization of the spatial profile of $J_u(v)$ in the vicinity of $v=\pm L_{\rm p}/4$. Note that the tape surface at $v=\pm L_{\rm p}/4$ is orientated parallel to the external field ($||~\hat{\bm{y}}$). We consider the non-monotonic $\beta$ dependence of $p(v)/\beta$ at the slots to be the origin of the functional form of $P_{\rm couple}(\beta)/\beta$ [Fig.~\ref{fig3}(a)]. The curve of $P_{\rm couple}(\beta)/\beta$ resembles the dependence of the coupling loss on the angular frequency of an alternating field \cite{campbell82}, but $P_{\rm couple}(\beta)/\beta$ scales as $1/\sqrt{\beta}$ at high $\beta$ for both $L_{\rm p}=2$~m and 6~m.

We obtain an approximate analytical formula for $P_{\rm couple}(\beta)/\beta$ at $\beta \ll \beta_{\rm c}$ through analysis in the thin-filament limit. In that limit, a multi-filament SC tape with resistive slots can be viewed as a continuous medium with the resistivity tensor
$\tilde{\rho}={\rm diag}[\rho_{||}(1+k^2u^2),\rho_{\perp}]$, where $\rho_\parallel$ (resp.\ $\rho_\perp$) is the longitudinal (resp.\ transverse) resistivity. The electromagnetic response of the continuous medium in the steady state is governed by the reduced Faraday--Maxwell equation
\begin{equation}
\frac{\partial}{\partial u}\left[ \rho_{||}\frac{\partial g}{\partial u}(1+k^2u^2) \right]+\frac{\partial}{\partial v}\left[\rho_{\perp}\frac{\partial g}{\partial v} \right]=\beta \cos kv.
\label{continuum-Faraday}
\end{equation}
For $\beta \ll \beta_{\rm c}$, the SC filaments are not coupled electromagnetically, meaning that the second term in Eq.~(\ref{continuum-Faraday}) is dominant over the first term, and therefore the first term can be neglected. By imposing the boundary condition $g(v=\pm L_{\rm p}/4)=0$, we obtain the solution $g(v)$, from which the transverse current density profile along the resistive slot at $u=0$ at low $\beta$ is
\begin{equation}
J_u(v)=-\frac{{\rm d}g(v)}{{\rm d}v}=-\frac{\beta}{k\rho_{\perp}}\sin kv.
\label{transverse-current-profile}
\end{equation}
We evaluate $\rho_{\perp}$ by means of the effective transverse resistance $R_{\perp}$,
\begin{eqnarray}
R_{\perp}
&=R^{\perp}_{\rm sc}+R_{\rm n}
=\frac{\rho_{\rm sc}w_1}{(L_{\rm p}/2)d_0}+\frac{\rho_{\rm n}s_1}{(L_{\rm p}/2)d_0}
=\frac{\rho_\perp w_1}{(L_{\rm p}/2)d_0},\nonumber\\
\end{eqnarray}
where $\rho_\perp\equiv\rho_{\rm sc}+(s_1/w_1)\rho_{\rm n}\approx (s_1/w_1)\rho_{\rm n}$. $R^{\perp}_{\rm sc}$ ($R_{\rm n}$) is the transverse SC (normal) resistance. We confirmed that the analytical formula for the transverse current density profile [Eq.~(\ref{transverse-current-profile})] agrees quantitatively with the numerical results for $J_u(v)$ along the resistive slot at $u=0$. In the case of four filaments, $P_{\rm couple}$ can be evaluated via $J_u$ as a quantitatively good approximation
\begin{eqnarray}
P_{\rm couple}(\beta)/\beta
&=\frac{d_0}{L_{\rm p}/2}\int\int_{\rm slots}{\rm d}u{\rm d}v(\rho_{\rm n}J_u^2+\rho_{\rm n}J_v^2)/\beta\nonumber \\
&\approx \frac{d_0\rho_{\rm n}}{L_{\rm p}/2}\int\int_{\rm slots}{\rm d}u{\rm d}v\left( -\frac{{\rm d}g(v)}{{\rm d}v}\right)^2/\beta\nonumber,\\
&=\frac{3d_0w^2_1L^2_{\rm p}}{8\pi^2s_1 \rho_{\rm n}}\beta.
\label{coupling-loss-low-sweep-rates}
\end{eqnarray}
Figure~\ref{fig5} shows that the analytical form of the coupling-loss power (\ref{coupling-loss-low-sweep-rates}) at low $\beta$ agrees quantitatively with the numerical results for $P_{\rm couple}(\beta)/\beta$.

At high $\beta$, the longitudinal SC resistance $R^{||}_{\rm sc}$ is three orders of magnitude greater than $R_{\rm n}$ at $\beta=1,000$~mT/s. Thus, the first term in Eq.~(\ref{continuum-Faraday}) is dominant over the second one. However, a quantitative analytical estimate of $P_{\rm couple}/\beta$ at high $\beta$ proved difficult because of the localization of $J_u(v)$ at the slots in the vicinity of $v=\pm L_{\rm p}/4$ due to the screening for an applied field at high $\beta$.


\section{Discussion}

We discuss the different mechanisms for electromagnetic coupling in an alternating magnetic field \cite{amemiya2004,amemiya2018} and a swept magnetic field.

In the case of an alternating field with angular frequency $\omega$, the electric current is time dependent, making the SC inductance crucial for the filament coupling criterion. The condition for the SC filaments to be electromagnetically decoupled is determined by the ratio of the SC inductive reactance $\omega L_{\rm sc}$ to the transverse normal resistance $R_{\rm n}$ across the slots, where $L_{\rm sc}=\mu_0 H (L_{\rm p}/2)/J_{\rm c}d_0$ the self-inductance of an SC filament and $\mu_0$ is the vacuum permeability. Thus, the SC filaments are electromagnetically decoupled when
\begin{eqnarray}
\frac{\omega L_{\rm sc}}{R_{\rm n}}&=\frac{\omega}{\omega_{\rm c}}\ll 1,
\end{eqnarray}
where the characteristic scale of the angular frequency for electromagnetic coupling is estimated as
\begin{eqnarray}
\omega_{\rm c}&\equiv \frac{R_{\rm n}}{L_{\rm sc}}\sim \frac{\rho_{\rm n}}{\mu_0 L^2_{\rm p}}\left( \frac{s_1}{d_0}\right).
\label{omega_c}
\end{eqnarray}
We evaluate the magnetic field approximately as $H\sim J_{\rm c}d_0/\pi$ assuming the Bean model.

In the steady state of a constantly ramped magnetic field, the electric current is time independent. In this case, the SC resistance is crucial for the filament coupling criterion, unlike in the case of an alternating field. Thus, the condition for the SC filaments to be electromagnetically decoupled is determined by the ratio of $R^{||}_{\rm sc}=\rho_{\rm sc}(L_{\rm p}/2)/w_1 d_0$ to $R_{\rm n}$. The maximum resistivity of the SC filaments is evaluated as $\rho_{\rm sc}\approx w_1\beta \cos kv/J_{\rm c}$ in the loosely twisted limit \cite{higashi2018}. Thus, the SC filaments are electromagnetically decoupled when
\begin{eqnarray}
\frac{R^{||}_{\rm sc}}{R_{\rm n}}&=\frac{\beta}{\beta_{\rm c}}\ll1,
\end{eqnarray}
where the characteristic scale of $\beta$ for electromagnetic coupling is estimated as
\begin{eqnarray}
\beta_{\rm c} &\sim \frac{J_{\rm c}\rho_{\rm n}s_1}{L^2_{\rm p}}.
\label{beta_c}
\end{eqnarray}

Here we note that an estimate is given for the critical value of the sample length for a slab geometry under an arbitrary time-varying applied magnetic field $B_{\rm a}(t)$ with a rate ${\rm d} B_{\rm a}(t)/{\rm d} t$ \cite{wilson1983}.
By setting $\beta={\rm d}B_{\rm a}/{\rm d}t= B_{\rm m}\omega$ for a fixed maximum amplitude of an alternating field $B_{\rm m}$,
we obtain the same results as Eqs.~(\ref{omega_c}) and (\ref{beta_c}) from the formula presented in Refs.~\cite{wilson1983,amemiya2004}
but the replacement of the length scale $s_1$ in Eqs.~(\ref{omega_c}) and (\ref{beta_c}) with $w_1$.
Thus, the characteristic scales, $\omega_{\rm c}$ and $\beta_{\rm c}$ are essentially the same results as those in Ref.~\cite{wilson1983},
but we stress that the origin determining the characteristic scale of the frequency and the sweep rate of an applied field is totally different as discussed the above.

We reason that the mechanism for electromagnetic coupling in a swept field differs from that in an alternating field, although the behavior of $P_{\rm couple}(\beta)/\beta$ (Fig.~\ref{fig3}) closely resembles the $\omega$ dependence of the coupling loss per cycle in an alternating magnetic field with a fixed field amplitude \cite{amemiya2006, comment}. Because the present model focuses on the steady state of a ramped field, transient behavior such as the coupling current between the SC filaments decaying with time cannot be taken account. That is, there is no characteristic time scale for the electromagnetic coupling in the steady state as there is in an alternating field. Nevertheless, $P_{\rm couple}(\beta)/\beta$ (Fig.~\ref{fig3}) exhibits a similar functional form to that for an alternating field because of the non-monotonic $\beta$ dependence of the spatial profile of the coupling-loss power density at the slots.

Next, we discuss the magnetization loss of a twisted stacked-tape cable (TSTC) conductor at high magnetic fields and comment on the possible application of twisted multi-filament tapes to dc magnets including an MRI magnet.
A single twisted tape itself has low current-carrying capacity, but it is improved by the TSTC technology \cite{takayasu2012}.
In this study, we are not modeling a TSTC conductor but a single twisted tape, and therefore the magnetic field shielding by the neighboring tapes is absent in the model,
although such an effect should be important \cite{grilli2006,zhang2017} to model a TSTC at low fields.
We note that even in the case of a TSTC conductor we can disregard the magnetic shielding effect.
This is because the fully penetrated state is realized for an applied field ($\sim$ a few tesla) which is above the full flux penetration field $B_{\rm p}$.
$B_{\rm p}$ for SC filaments is approximately estimated by the formula for a flat tape as $B_{\rm p}=(\mu_0 J_{\rm c}d_0/\pi)[1+\ln(w_1/d_0)]\approx0.287$ T.
In the case of stacked tapes, $B_{\rm p}$ increases several times larger than $B_{\rm p}$ for a single tape, but magnetic fluxes penetrate fully into stacked tapes at high fields.
Thus, the magnetization loss of a TSTC conductor is supposed to be the sum of the loss on each twisted tape,
and therefore the single tape model in this article would be useful for the analysis of magnetization losses of a multi-filament TSTC conductor at high fields.

Last of all, we comment on the possible application to dc magnets such as an MRI magnet.
The condition that electromagnetic coupling is suppressed for the field sweep rate of a practical MRI magnet is restricting $L_{\rm p}$ to no more than a few meters.
Successful suppression of electromagnetic coupling can be achieved for one-turn superconductor coil with a bore diameter of $\sim 1$ m,
composed of loosely twisted stacked multi-filament tapes with $w_0/L_{\rm p} \sim 0.002$
if each tape is insulated.
Mechanical strain due to the twist along the stack axis can be mostly avoided because of the loosely turned coil structure.
Indeed, the 50-tape, 2.5-turn superconductor test coil
composed of ReBCO tapes wound on a pentagon cylinder with the diameter of 165 mm has been fabricated \cite{takayasu2013}.
It is possibly applicable to high energy physics magnets or fusion magnets \cite{takayasu2013,takayasu2016,takayasu2017}.

\section{Summary}
We performed numerical simulations on the magnetization loss on a twisted multi-filament SC tape in the steady state on the basis of a thin-sheet approximation. The dependence of the power loss on the field sweep rate $\beta$ shows the absence of electromagnetic coupling of the SC filaments for $\beta \ll \beta_{\rm c}$ (coupling sweep rate), making it possible to reduce the power loss by multi-filamentarization. For $\beta \geq \beta_{\rm c}$ (i.e., $R^{||}_{\rm sc} \ge R_{\rm n}$), the filaments are coupled electromagnetically, making multi-filamentarization ineffective.
Because $\beta_{\rm c} \propto 1/L^2_{\rm p}$, restricting $L_{\rm p}$ to no more than a few meters makes it possible to make $\beta_{\rm c}$ sufficiently large compared with the typical sweep rate of $\sim\!1$~mT/s of an MRI magnet.

The $\beta$ dependence of the coupling-loss power per unit $\beta$ exhibits a similar functional form to that in an alternating field. However, the mechanism for electromagnetic coupling differs from that in an alternating field because $\beta_{\rm c}$ is determined by the ratio of the longitudinal SC resistance to the normal resistance of the slots in a swept field.

\ack
We would like to thank our colleagues at AIST, namely, T.~Izumi, T.~Machi, M.~Furuse, H.~Takashima, S.~Ishida, and Y.~Yoshida, for discussions and valuable comments. This work is based on results obtained in a project commissioned by the New Energy and Industrial Technology Development Organization (NEDO).


\section*{Appendix}
In the coordinates $(u,\eta,v)$ of the twisted SC tape, an arbitrary vector $\bm{O}=(O_x,O_y,O_z)$ is expressed as 
$\bm{O}=O_u\hat{\bm{u}}+O_\eta\hat{\bm{\eta}}+O_v\hat{\bm{v}}$ with
$\hat{\bm{u}}=\hat{\bm{x}}\cos kv+\hat{\bm{y}}\sin kv$,
$\hat{\bm{\eta}}=\hat{\bm{x}}\sin kv + \hat{\bm{y}}\cos kv$, and
$\hat{\bm{v}}=(ku\hat{\bm{\eta}}-k\eta\hat{\bm{u}}+\hat{\bm{z}})/\sqrt{1+k^2(u^2+\eta^2)}$.
Thus,
\begin{eqnarray}
O_x&=
O_u\cos kv -O_\eta \sin kv \nonumber \\
&-O_v k(u \sin kv + \eta \cos kv)/\sqrt{1+k^2(u^2+\eta^2)},
\label{Ox}\\
O_y&=
O_u \sin kv + O_\eta \cos kv\nonumber \\
&+O_v k(u \cos kv -\eta \sin kv)/\sqrt{1+k^2(u^2+\eta^2)},
\label{Oy}\\
O_z&=O_v/\sqrt{1+k^2(u^2+\eta^2)}.
\label{Oz}
\end{eqnarray}
On the tape surface ($\eta=0$), we have from Eqs.~(\ref{Ox})--(\ref{Oz}) that $O^2_x+O^2_y+O^2_z=O^2_u+O^2_v$.
\section*{References}
\bibliography{iopart-num}

\end{document}